# A Synergistic Approach: Dynamics-AI Ensemble in Tropical Cyclone Forecasting

Yonghui LI[1,2,3], Wansuo DUAN*[1,2], Hao LI*[4,3], Wei HAN[5,6], Han ZHANG[7] and Yinuo LI[8]

[1]*State Key Laboratory of Earth System Numerical Modeling and Application, Institute of Atmospheric Physics, Chinese Academy of Sciences, Beijing, 100029, China*

[2]*University of Chinese Academy of Sciences, Beijing, 101408, China*

[3]*Shanghai Academy of Artificial Intelligence for Science (SAIS), Shanghai, 200232, China*

[4]*Artificial Intelligence lnnovation and Incubation Institute, Fudan University, Shanghai, 201203, China*

[5]*State Key Laboratory of Severe Weather Meteorological Science and Technology (LASW), Beijing, 100081, China*

[6]*CMA Earth System Modeling and Prediction Centre (CEMC), China Meteorological Administration, Beijing, 100081, China*

[7]*Henan Meteorological Observatory, Zhengzhou, 450003, China*

[8]*Tianjin Meteorological Information Center, Tianjin, 300074, China*

## ABSTRACT

This study addresses a critical challenge in AI-based weather forecasting by developing an AI-driven optimized ensemble forecast system using Orthogonal Conditional Nonlinear Optimal Perturbations (O-CNOPs). The system bridges the gap between computational efficiency and dynamic consistency in tropical cyclone (TC) forecasting. Unlike conventional ensembles limited by computational costs or AI ensembles constrained by inadequate perturbation methods, O-CNOPs generate dynamically optimized perturbations that capture fast-growing errors of FuXi model while maintaining plausibility. The key innovation lies in producing orthogonal perturbations that respect FuXi's nonlinear dynamics, yielding structures reflecting dominant dynamical controls and physically interpretable probabilistic forecasts. Demonstrating superior deterministic and probabilistic skills over the operational Integrated Forecasting System Ensemble Prediction System, this work establishes a new paradigm combining AI's computational advantages with rigorous dynamical constraints. Success in TC track forecasting paves the way for reliable ensemble forecasts of other high-impact weather systems, marking a major step toward operational AI-based ensemble forecasting.

___________________________

*Corresponding author : Wansuo DUAN
Email: duanws@lasg.iap.ac.cn
*Corresponding author : Hao LI
Email: lihao_lh@fudan.edu.cn

**Article Highlights:**

- Developed an AI-driven ensemble system using O-CNOPs to generate physically consistent, optimized perturbations for TC forecasts
- O-CNOPs capture realistic forecast uncertainties for TC Tracks in FuXi model, enabling reliable forecasts
- The system generally outperforms operational forecasts in deterministic and probabilistic scores

**1. Introduction**

Tropical cyclones (TCs) are often accompanied by extreme weather events such as strong winds, torrential rains, massive waves, and storm surges, causing severe damage to life and property (Gori et al. 2022; Peduzzi et al. 2012; Zhang et al. 2017). It is therefore extremely important for disaster prevention and reduction to forecast accurately TCs. However, due to the highly chaotic nature of TC systems (Wang and Wu 2004; Nystrom et al. 2018), even minor errors in initial conditions can rapidly amplify over time, leading to significant forecast uncertainty (Zhang et al. 2023; Heming 2017; Li et al. 2025), as found in the Lorenz system (Lorenz 1963, 1969). To address this, forecasters typically use ensemble forecasting strategy, which generates multiple possible scenarios to quantify this uncertainty, thereby improving forecast reliability and providing the scientific basis for disaster prevention decision-making.

The ensemble strategy has substantially enhanced TC forecasting performance in major meteorological departments such as ECWMF, NECP, etc. (Mureau et al. 1993; Toth and Kalnay 1993; Houtekamer et al. 1996; Bourke et al. 1995), but they demand extensive computational resources due to reliance on complex physics-based numerical models. Nevertheless, it is very exciting that high-efficient artificial intelligence (AI) techniques have emerged in recent years as powerful tools for forecasting model development. Several AI-driven large meteorological models - including FuXi (Chen et al. 2023), Pangu-Weather (Bi et al. 2023), AIFS (Lang et al. 2024a), and GraphCast (Lam et al. 2023), etc. - have attracted considerable attention for their exceptional computational efficiency and predictive accuracy. These models present viable alternatives to traditional numerical approaches. Particularly, Li et al. (2025) adopted the FuXi model to explore TC track predictability and found that initial perturbations exhibit significant growth phenomenon; furthermore, Pu et al. (2025) revealed that initial perturbations of particular spatial structure enhance such error growth and are favorable for ensemble forecasts. That is to say, similar to the Lorenz system (Lorenz 1969, 1963), the output of the FuXi model exhibits extreme sensitivity on initial input data uncertainties. This discovery unlocks a transformative opportunity—by integrating dynamical ensemble methods with AI architectures like FuXi, one can achieve quantum leaps in both computational efficiency and predictive reliability, fundamentally overcoming the trade-offs of traditional approaches.

Currently, it is popular for ensemble forecasts that incorporate rapidly-growing initial perturbations with physics-informed models to generate ensemble members. The major methods to yield rapidly-growing initial perturbations are singular vectors (Buizza and Palmer 1995; Mureau et al. 1993) (SVs) and breeding vectors (Toth and Kalnay 1993) (BVs); however, these two methods either rely on linear dynamics and cannot depict nonlinear nature of atmospheric and oceanic motions or are uncertain to represent fast-growing perturbation during forecast period, indeed hindering the enhancement of ensemble forecast capability (Zhang et al. 2023; Duan et al. 2023). Regarding this limitation, a novel method of orthogonal conditional nonlinear optimal perturbations (Duan and Huo 2016) (O-CNOPs) has been proposed to generate the fast-growing initial perturbations for ensemble forecasts. This method fully accounts for nonlinear physical processes and has been shown to possess better performance than SVs and BVs in TC track forecasting (Zhang et al. 2023; Huo and Duan 2019). Related to the FuXi model, it is certainly concerned that whether its integration with O-CNOPs can additionally significantly enhance TC track ensemble forecasting level besides its high efficiency as expected.

In this study, we integrated the O-CNOPs with the FuXi model and established a novel ensemble forecasting system (hereafter as "FuXi_CNOP") to TC track forecasts. We revealed that, with only a small number of members produced, it can achieve markedly improved forecasting capability on TC track, even higher than the world leading ensemble forecast system. It is therefore that the study provides a highly promising approach to produce ensemble for AI model forecasts on TC track.

## 2. Method and Experiment Setup

### 2.1 Orthogonal Conditional Nonlinear Optimal Perturbations

The orthogonal conditional nonlinear optimal perturbations (O-CNOPs) (Duan and Huo 2016) are a group of mutually orthogonal initial perturbations with maximum growth in their respective subspaces (see Eq. (1)).

$$J\left(\delta \mathbf{x}_{0j}^*\right) = \max_{\delta \mathbf{x}_{0j} \in \mathbf{\Omega}_j} \| M_\tau\left(\mathbf{x}_0 + \delta \mathbf{x}_{0j}\right) - M_\tau(\mathbf{x}_0) \|_{C_2}, \tag{1}$$

where

$$\mathbf{\Omega}_j = \begin{cases} \delta \mathbf{x}_{0j}^* \in \mathbb{R}^n | \ \| \delta \mathbf{x}_0 \|_{C_1} \leq \epsilon & , j = 1 \\ \delta \mathbf{x}_{0j}^* \in \mathbb{R}^n | \ \| \delta \mathbf{x}_{0j} \|_{C_1} \leq \epsilon, \delta \mathbf{x}_{0j} \perp \mathbf{\Omega}_k, k = 1, \cdots, j-1, \ j > 1 \end{cases}. \tag{2}$$

Here, $M_\tau$ is a nonlinear operator that evolves a state from initial time $t = 0$ to a future forecast time $t = \tau$; $\mathbf{x}_0$ is an initial condition and $\delta \mathbf{x}_0$ is its initial perturbation; $\mathbb{R}^n$ denotes $n$-dimension Euclidean space; $\mathbf{\Omega}_j$ is one of subspaces in $\mathbb{R}^n$ and defines an initial perturbation space; the norms $\| \bullet \|_{C_1}$ and $\| \bullet \|_{C_2}$ are the measures of the initial perturbation and its evolution at forecast time, respectively; the symbol "$\perp$" signifies the

orthogonality of vector spaces. According to Eqs. (1) and (2), the 1st CNOP has the maximum nonlinear growth in the first subspace, i.e., the entire perturbation phase space $\mathbf{\Omega}_1$; and the $j$ th CNOP provides the maximum nonlinear growth in the subspace orthogonal to the first $j-1$ CNOPs. O-CNOPs can be seen as an extension of traditional SVs, employed in the IFS-EPS in ECMWF, in the nonlinear regime, fully taking into account the impact of nonlinear physical processes.

In the present study, the initial perturbation $\delta \mathbf{x}_0$ is composed of $\mathbf{u}_0^{'}$, $\mathbf{v}_0^{'}$, $\mathbf{T}_0^{'}$, which correspond to zonal and meridional winds and temperature, respectively. The norm $\parallel \bullet \parallel_{C_1}$ is taken as the sum of kinetic and internal energy (see Eq. (3)) and measures amplitude of initial perturbations, $\parallel \bullet \parallel_{C_2}$ measures the perturbation kinetic energy at the forecast time $\tau$ (see Eq. (4)). Specifically, they have the expressions as follows.

$$\parallel \delta \mathbf{x}_0 \parallel_{C_1}^2 = \frac{1}{D_1} \int_\sigma \int_{D_1} \left[ \mathbf{u}_0^{'2} + \mathbf{v}_0^{'2} + \frac{c_p}{T_r} \mathbf{T}_0^{'2} \right] dD_1 \, d\sigma \tag{3}$$

and

$$J = \parallel M_\tau(\mathbf{x}_0 + \delta \mathbf{x}_0) - M_\tau(\mathbf{x}_0) \parallel_{C_2} = \frac{1}{D_2} \int_\sigma \int_{D_2} \left[ \mathbf{u}_\tau^{'2} + \mathbf{v}_\tau^{'2} \right] dD_2 d\sigma \tag{4}$$

where $c_p$ is specific heat at a constant pressure with value of 1005.7 J kg⁻¹ K⁻¹, the reference parameter $T_r = 270 \, K$, $\mathbf{u}_0^{'}$ and $\mathbf{v}_0^{'}$ are the wind components of initial perturbation, $\mathbf{T}_0^{'}$ is the temperature component, $D_1$ covers the global atmosphere and $\sigma$ is the vertical coordinate from 13 pressure levels of FuXi model; $J$ is the objective function defined as the kinetic energy over the verification area $D_2$, which covers corresponding TC's initial position and entire track trajectory with uniform 5° expansion in all cardinal directions, in attempt to fully interpret the positional uncertainties of the TC circulation (Zhang et al. 2023); $\mathbf{u}_\tau'$ and $\mathbf{v}_\tau'$ are the evolution of the wind components of initial perturbation $\delta\mathbf{x}_0$ at the forecast time $\tau$.

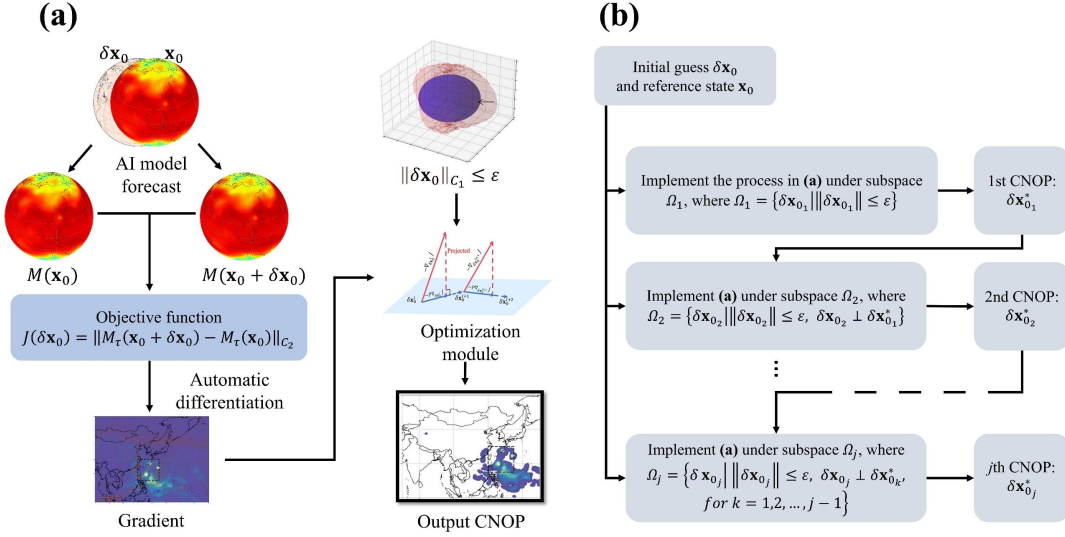

**Fig. 1.** Workflow of O-CNOPs computation. (a) The process searching for CNOP using the automatic differentiation module in an AI model. (b) The flow of O-CNOPs computatiom.

Applying the Eqs. (3) and (4) to the Eq. (1), one can calculate the first CNOP through automatic differentiation module robustly existed in AI models—a process that implements objective function using the PyTorch's computational graph (Paszke et al. 2017) and utilizes automatic differentiation module to calculate the gradient of the objective function (Li et al. 2025) (see Fig. 1a). Then, in the subspace orthogonal to the first CNOP $\delta\mathbf{x}_{01}^*$, one computes the second CNOP $\delta\mathbf{x}_{02}^*$, in the subspace orthogonal to both the first and second CNOPs, computes the third CNOP $\delta\mathbf{x}_{03}^*$, ..., in the subspace orthogonal to the first $j-1$ CNOPs, one can calculate the $j^{\text{th}}$ CNOP $\delta\mathbf{x}_{0j}^*$ (see Fig. 1b). Then a group of O-CNOPs can be achieved for ensemble forecasts.

## 2.2 FuXi Model

The FuXi model (Chen et al. 2023) was developed by the Artificial Intelligence Innovation and Incubation Institute at Fudan University. It employs a cascaded neural architecture to generate 15-day global atmosphere forecasts at 0.25° spatial and 6-hour temporal resolutions. It incorporates 70 state variables, utilizing the input data combining both current and previous time steps with dimensions of $2 \times 70 \times 721 \times 1440$. This model is trained utilizing a 36-year (1979-2015) ERA5 reanalysis dataset. Chen et al. (Chen et al. 2023) demonstrated that the forecast performance, evaluated using latitude-weighted root mean square error and anomaly correlation coefficient, is comparable to the ECMWF ensemble mean even at a 15-day lead time, making it the first AI-based model to achieve such accuracy.

### *2.3 Experimental strategy*

The ensemble forecasting skill is sensitive to ensemble parameters including amplitudes of initial perturbations, optimization time interval and ensemble size $N$ (Duan et al. 2023; Toth and Kalnay 1997). For the FuXi_CNOP as stated in the context, it, based on strategical sensitivity experiments (see Fig. S1 in Supplementary; also referred to Zhang et al. (2023)), determines the ensemble parameters as perturbation amplitude $\epsilon = 0.21 \, \mathrm{J \, kg^{-1}}$, optimization time interval $\tau = 0.5$ days, and ensemble size 31, which comprise one control forecast and 30 ensemble members produced by superimposing 15 pairs of positive and negative CNOP onto the control forecast. The control forecasts are generated by the FuXi model, which are initialized by 12-hour lead time forecast fields starting from ERA5 reanalysis data.

The study evaluated TC track ensemble forecasts generated by FuXi_CNOP as compared with the Integrated Forecast System Ensemble Prediction System (IFS_EPS) of ECMWF, the current benchmark for numerical weather prediction-based ensemble forecasting (Conroy et al. 2023). We utilized the TIGGE dataset containing ECMWF's IFS version 48r1 forecasts (native 9 km/137 levels, interpolated to 0.25° resolution). The TC best track dataset used is the IBTrACS dataset. A comprehensive description of the TC tracker can be referred to Zhong et al. (2023). This comparative framework enables rigorous assessment of both deterministic and probabilistic forecast skills while accounting for realistic initialization scenarios.

We conducted comprehensive diagnostic comparisons between FuXi_CNOP and IFS_EPS. The analysis focuses on two primary TC basins: the Western North Pacific and North Atlantic. To ensure rigorous out-of-sample validation, our analysis focused exclusively on TC events occurring during 2018-2023 - a period completely independent from FuXi's training period of 1979-2015. The study assessed TC track forecast skill through 91 forecasts of 62 TCs, including 51 cases of 35 TCs over the Western North Pacific and 40 cases of TCs over the North Atlantic, all of whose intensities reach tropical storm criteria. This evaluation framework enables direct comparison between the FuXi_CNOP and IFS_EPS under identical verification conditions.

## 3. Results

### 3.1 *Comparative analysis of FuXi_CNOP and IFS_EPS in TC Deterministic Forecasting*

In this section, we make a comprehensive comparison between IFS_EPS and our newly developed FuXi_CNOP, as based on analysis of a total of 91 cases including 51 cases occurring in Northwest Pacific and 40 cases arising in the North Atlantic. Note that the IFS_EPS employs singular vectors and ensemble data assimilation to generate initial perturbations, which combined with stochastically physics perturbations, to account for both initial and model uncertainties of control forecasts generated by the IFS, with 51 ensemble members; however, our FuXi_CNOP only generates initial perturbations using the O-CNOPs method and specifically designed to estimate initial uncertainties through 31 ensemble numbers.

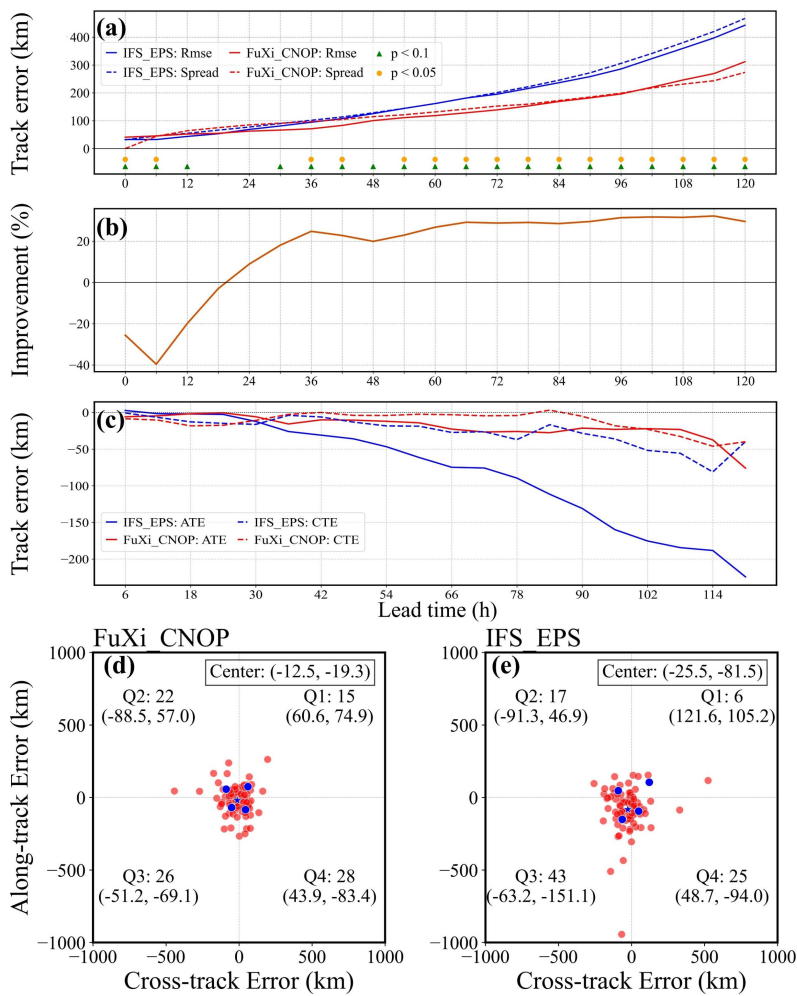

**Fig. 2**. Evaluation of deterministic forecasting skills. (a) Ensemble mean track errors (solid lines) and spread (dashed lines) for FuXi_CNOP (red) and IFS_EPS (blue) across 91 cases. Yellow circles and green triangles indicate statistically significant differences at 95% and 90% confidence levels, respectively (Student's t-test). (b) Percentage reduction in ensemble mean track forecast errors of FuXi_CNOP relative to IFS_EPS. (c) Mean along-track errors (ATE, solid lines) and cross-track errors (CTE, dashed lines) as a function of lead time, averaged across 91 cases. (d) Scatter distribution of ATE vs. CTE for FuXi_CNOP, with quadrant means (blue dots) and overall means (blue stars) shown for reference. (e) Same as (d) but for IFS_EPS.

Fig. 2 presents a comparative evaluation of the FuXi_CNOP and IFS_EPS in TC track forecasting, focusing on ensemble mean forecast errors, ensemble spread, along-track errors (ATE), and cross-track errors (CTE) (see Appendix). In the short lead-time (0–12 h) forecasts, both systems exhibit comparable performance. However, beyond 18 h, the FuXi_CNOP demonstrates a significant reduction in forecast errors, significantly outperforming the IFS_EPS from 24 h onward and maintaining this advantage up to 120 h. This improvement is statistically significant (p<0.1) between 36–120 h, with a maximum error reduction of 32.33%. Notably, the FuXi_CNOP achieves a near-unity skill-spread ratio, mirroring the reliability of IFS_EPS in characterizing forecast uncertainty. This dual capability—enhanced accuracy and robust uncertainty quantification—highlights its superiority in long-range TC track forecasts.

In addition, the FuXi_CNOP demonstrates substantially lower ATE and CTE compared to the IFS_EPS beyond the 24-hour lead time. Especially, the FuXi_CNOP maintains near-zero ATE values, reflecting precise translation speed estimation, whereas the IFS_EPS exhibits persistent negative error with indicative of systematic underestimation of TC motion velocity. In CTE performance, the FuXi_CNOP outperforms the IFS_EPS with an indication of significantly reduced leftward track biases. The scatter distributions of ATE-CTE phase points for all 91 cases (see Fig. 2d and Fig. 2e) reveal two distinct patterns: the FuXi_CNOP's phase points cluster tightly around the coordinate origin, with the one averaged for 91 cases positioned close to the coordinate origin, confirming trivially-unbiased performance in TC motion; in stark contrast, the IFS_EPS phase points mostly concentrate in the third quadrant, with the one averaged for all TCs displaced accordingly - a spatial signature of combined slow-speed bias and consistent leftward track deviations in predicted TC tracks.

It is obvious that, for the deterministic forecasts in terms of the ensemble mean, ensemble spread, ATE and CTE, the FuXi_CNOP possesses aggressively more advantages than the IFS_EPS in not only forecasting accuracy but also performance consistency for TC tracks. The study further categorized TCs into two distinct basins, i.e., above-mentioned the Western North Pacific (51 cases) and the North Atlantic (40 cases), with consistent results observed (see Fig. S2 and Fig. S3).

### 3.2 *Comparative analysis on Uncertainty Quantification Provided by FuXi_CNOP and IFS_EPS in TC Track Probability Forecasting*

The Continuous Ranked Probability Score (CRPS) is a rigorously defined statistical metric for quantifying the accuracy of probabilistic forecasts by measuring the disparity between a predicted cumulative distribution function (CDF) and the observed value of a continuous variable (see Appendix). Lower CRPS values denote superior performance. In this study, we apply CRPS to assess the probabilistic forecasting skill of TC track predictions by comparing model outputs against best-track observations. Fig. 3 presents comparative CRPS results for FuXi_CNOP and IFS_EPS across forecast lead times. It shows that, for the lead times 0-12 hours, the two systems demonstrate comparable CRPS values, while beyond 18 hours, the FuXi_CNOP shows obvious advantages, especially from 24 hours overtaking the IFS_EPS and through 120 hours maintaining consistent superiority (see Fig. 3a), achieving a maximum CRPS reduction of 29.2% (see Fig. 3b).

These advantages for the FuXi_CNOP are statistically significant (p<0.1) for 36-120-hour lead times, with most lead times reaching p<0.05. These findings demonstrate that FuXi_CNOP not only improves the accuracy of ensemble-mean track forecasts but also provides a reliable representation of forecast uncertainty (i.e. the first and second moments of the probability density distribution), thereby delivering a more skillful probabilistic characterization of TC track evolution at extended lead times. The study also categorized TCs into the Western North Pacific and the North Atlantic and consistent results were obtained (see Fig. S4 and Fig. S5).

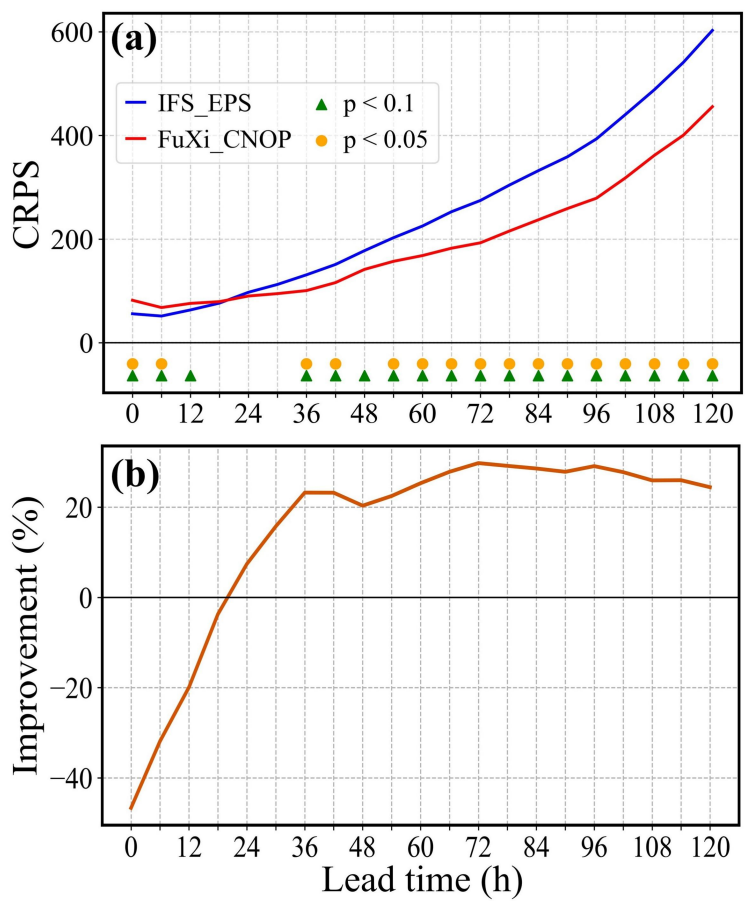

**Fig. 3.** Evaluation of CRPS. (a) CRPS comparison for 91 cases between FuXi_CNOP (red) and IFS_EPS (blue). Yellow circles (95% confidence) and green triangles (90% confidence) indicate

statistically significant differences (Student's t-test). (b) Percentage improvement of FuXi_CNOP CRPS relative to IFS_EPS.

More specifically, the study explored the TC track strike probability in ensemble forecasts generated by FuXi_CNOP and IFS_EPS, where both Brier Score (BS) and Receiver Operating Characteristic (ROC) curve are adopted to measure the probability. We find that the FuXi_CNOP achieves BS scores comparable to those of the IFS_EPS (see Fig. S6), while its ROCA performance is marginally inferior to the IFS_EPS (see Fig. S7). This indicates that the FuXi_CNOP achieves comparable reliability to the IFS_EPS in probabilistic strike forecasts, as reflected by similar BS values, but its slightly inferior ROC performance highlights a trivially reduced discrimination capability. In other words, while FuXi_CNOP tends to assign probabilities that match IFS_EPS' strike frequencies (see Fig. 4 for some examples), its ability to more clearly separate strike from non-strike cases is somewhat weaker than IFS_EPS, resulting in less distinction between high- and low-risk scenarios.

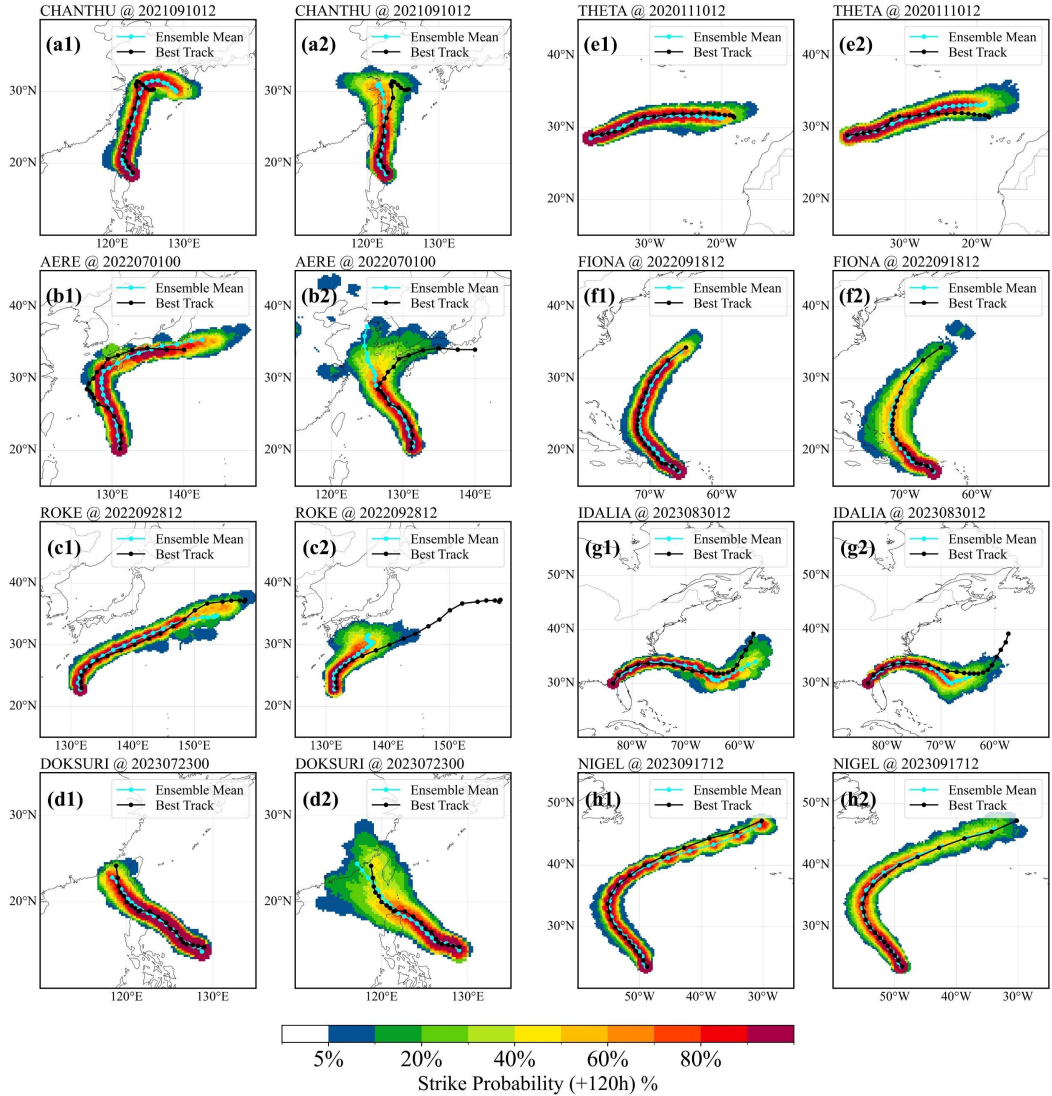

**Fig. 4**. TC strike probability forecasts for TCs (a-d) CHANTHU, AERE, ROKE, and DOKSURI in the Northwest Pacific, and (e-h) THETA, FIONA, IDALIA, and NIGEL in the North Atlantic. The suffix '1' denotes FuXi_CNOP, while '2' represents IFS_EPS. Forecasts include the best track (black line), ensemble mean (blue line), and strike probability density (shaded).

In summary, the FuXi_CNOP demonstrates significantly superior performance compared to the IFS_EPS in deterministic forecasts across four key metrics: ensemble mean accuracy, ensemble spread consistency, ATE, and CTE. While the two systems exhibit comparable reliability in TC track strike probability forecasts—with FuXi_CNOP showing marginally lower discrimination ability—the former achieves more skillful probabilistic characterization of TC track evolution at extended lead times. Consequently, the FuXi_CNOP generally outperforms the IFS_EPS in both deterministic and probabilistic TC track forecasting, especially for long-range lead times.

### 3.3 *Mechanism*

Building on the results of deterministic and probabilistic forecasts of the FuXi_CNOP with respect to the TC tracks, this section elucidates the underlying mechanism that enables superior forecast skill of TC tracks. By observing the first three CNOPs of each TC, it is found that their horizontal structures reveal their exceptional capability in identifying key sensitive areas responsible for TC track forecasting uncertainty. Fig. 5 shows representative examples of the CNOPs for TC CHANTHU in the Northwest Pacific and THETA in the North Atlantic. For the TC CHANTHU, its perturbation energies characterized by the CNOPs are found to predominantly concentrate in the spiral rainbands of the TC (see Fig. 5a), which often directly interface with the subtropical high periphery, consequently determining the TC track; another case for TC THETA in North Atlantic (see Fig. 5b), its perturbation energies display a dual-core structure, with particularly pronounced perturbation maxima: the primary maximum located within the southern subtropical high interaction zone, which shows strong dynamical coupling with the TC's subsequent northward track deflection, and the secondary maximum situated in the southwestern eyewall region, with the asymmetry of the eyewall modulating the evolution of the TC wind field and, in turn, exerting an influence on the TC track[24]. It is obvious that the CNOP patterns identify the dynamical instability within the TC region that captures the environmental modulation of TC circulation fields, consequently enhancing subsequent TC track forecast skill-a finding consistent with previous studies (Zhang et al. 2023; Zhou and Mu 2011).

Now we adopt the statistics of 91 cases to elucidate the vertical energy distributions of the CNOPs (see Fig. 5c). It shows that the kinetic energy generally peaks at 400-300 hPa, coinciding with the TC warm-core structure (Li et al. 2025) and reflecting crucial baroclinic energy conversion processes (see Fig. S8), while internal energy maximizes in the 925-700 hPa boundary layer, corresponding to moisture transport (Li and Wang 2021) and latent heat release mechanisms.The superimposed profile in Fig. 5c reveals a dual-peak vertical structure in energy distribution, demonstrating that the FuXi_CNOP model successfully captures both mid-upper level baroclinic processes and lower-level moist convective processes—the two primary physical mechanisms governing TC track uncertainty. Furthermore, the vertical energy distribution patterns of CNOPs for TCs in the Western North Pacific and North Atlantic basins exhibit remarkable similarity (see Fig. S9), highlighting consistent dynamical behaviors across these geographically distinct regions.

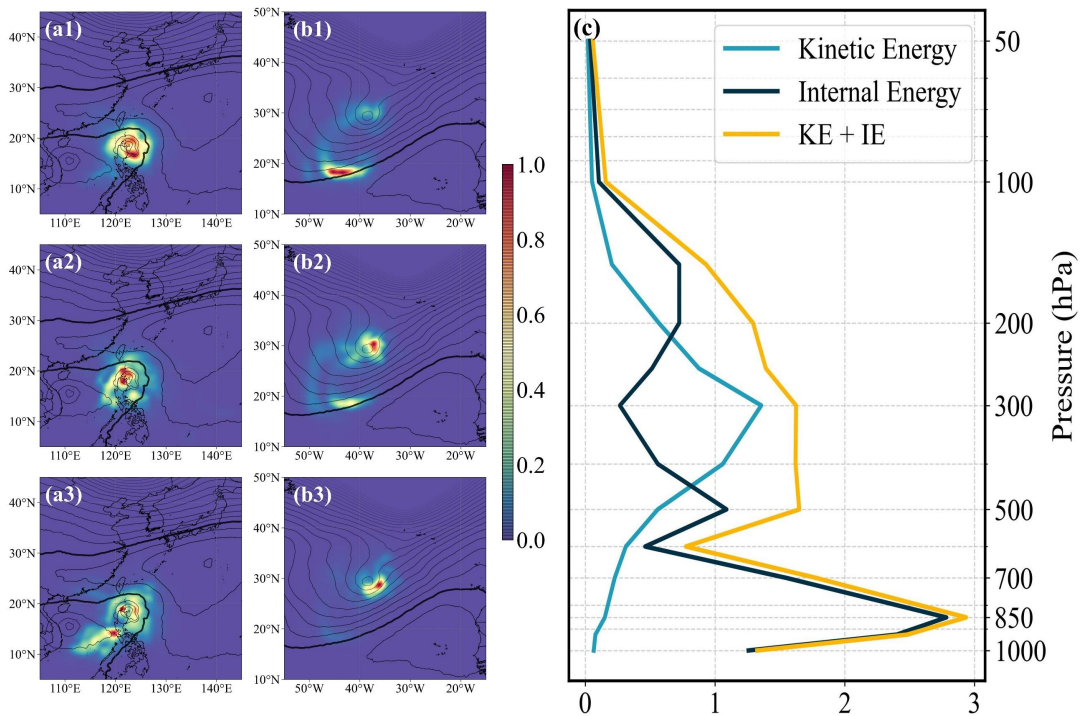

**Fig. 5**. Horizontal and vertical perturbation structures of CNOPs. (a) TC CHANTHU (09/10/2021 12Z) in the Northwest Pacific and (b) TC THETA (11/10/2020 12Z) in the North Atlantic, showing vertically integrated CNOP energy (shaded, normalized 0–1) with geopotential height contours (black lines; thick line indicates 5880 gpm). Labels 1, 2, and 3 denote the 1st, 2nd, and 3rd CNOP pattern, respectively. (c) Mean horizontally integrated energy profiles for 91 cases: kinetic energy (sky blue), internal energy (dark blue), and their sum (yellow).

As indicated by the CNOP structures, the CNOPs effectively identify the key areas responsible for TC track forecasting uncertainties and capture both mid-upper level baroclinic processes and lower-level moist convective processes governing TC track uncertainty. We therefore have strong reason to believe that the ensemble members generated by the FuXi_CNOP are capable of representing TC track forecasting uncertainty and provide reliable forecast of TC tracks. It is known that TC motion is primarily modulated by the steering flow related to subtropical high pressure systems (Zhang et al. 2023; Kossin et al. 2010; Wu et al. 2005). We plot in Fig. 6 the ensemble spread of the wind generated by the FuXi_CNOP and IFS_EPS for the TCs as in Fig. 4. It is shown that FuXi_CNOP demonstrates superior dynamical consistency in representing the steering flow interactions, as evidenced by the coherent spatial coupling between its 850-250 hPa mean wind fields and wind speed spread pattern. Specifically for the TC cases of CHANTHU and THETA (see Fig. 6a1 and 6e1; consistent with the case selections in Fig. 5), it is illustrated that the wind speed spread exhibits a clear correspondence with the sensitive areas and is well coupled with the TC steering flow, demonstrating that the CNOPs effectively captures the uncertainties that influence TC track evolution. Further, we adopt the TC AERE to exemplify this advantage (see Fig. 6b1): the system not only accurately resolves the TC core structure, known to critically influence TC development (Chen et al. 2018; Qin et al. 2020), but also optimally localizes maximum spread regions along the shear zone between the southwesterly steering flow and TC circulation - precisely corresponding to the primary sources of track forecast uncertainty; in contrast, the IFS_EPS exhibits spread maxima not only in the TC vicinity but also prominently in the South China Sea monsoon region (15°N-20°N, 110°E-120°E) (see Fig. 6b2). This dual-maxima pattern indicates that the system's uncertainty representation is not optimally focused on TC-specific dynamics, instead maintaining broadly distributed spread across multiple synoptic systems. These findings demonstrate that the FuXi_CNOP enable more precise identification and representation of key uncertainty sources in TC track forecasting, thus providing an interpretation on the superior performance of FuXi_CNOP in TC track forecasting.

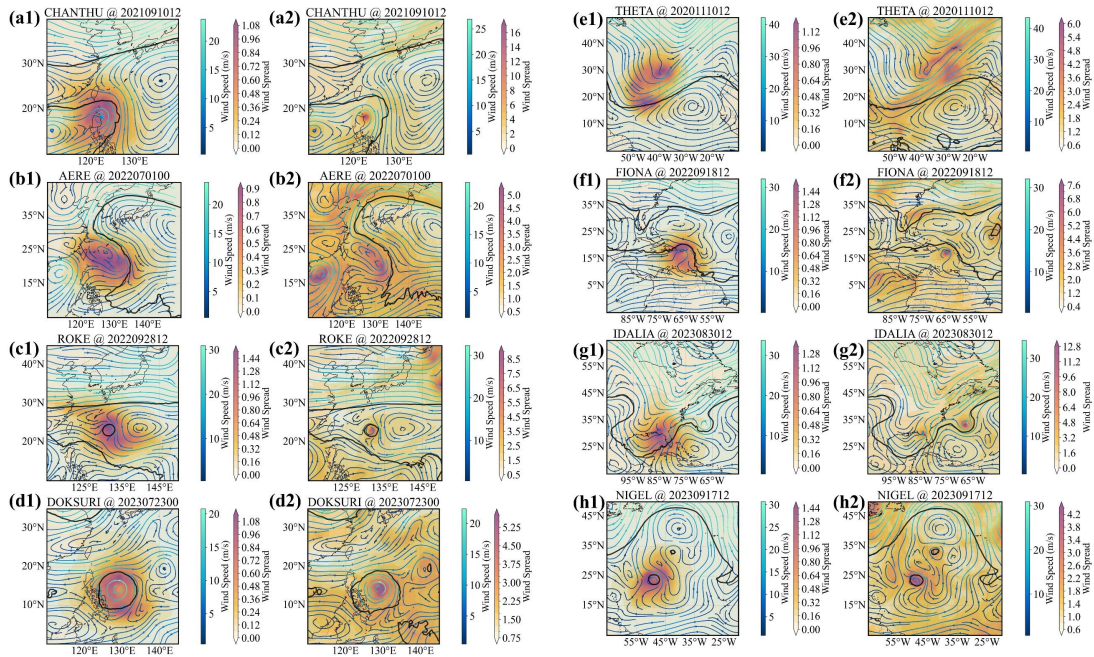

**Fig. 6**. Ensemble-mean deep-layer (850-250 hPa) wind fields (streamlines) and associated wind speed spread (shading) at initial time. (a-d) CHANTHU, AERE, ROKE, and DOKSURI in the Northwest Pacific, and (e-h) THETA, FIONA, IDALIA, and NIGEL in the North Atlantic, with suffix "1" denoting FuXi_CNOP results and "2" representing IFS_EPS outputs. The 5880 gpm geopotential height contour at 500 hPa is overlaid in black.

## 4. Dicussion

This study introduces a novel AI-driven ensemble forecasting framework that synergies dynamical O-CNOPs optimization with advanced meteorological modeling. By integrating the FuXi AI large model, the system achieves unprecedented computational efficiency while maintaining rigorous dynamical consistency. When specifically applied to TC track forecasting, the FuXi_CNOP framework demonstrates comprehensive superiority over the benchmark IFS_EPS system: it exhibits enhanced ensemble mean accuracy, superior ensemble spread consistency, and significant reductions in both along-track and cross-track errors. While both systems provide comparable reliability in probabilistic strike forecasts, FuXi_CNOP establishes a distinct advantage in deterministic track forecast and generally excels in probabilistic characterization of TC evolution at extended lead times. Physically, the framework's perturbation energy structures reveal distinct spatial patterns: horizontal energy concentrates in TC core regions and subtropical high interface zones, while vertical energy shows dual-peak distributions corresponding to baroclinic and moist convective processes; these structural characteristics enable FuXi_CNOP to more effectively capture critical environmental factors governing TC motion, particularly through improved representation of subtropical high-TC circulation interactions; consequently, the framework delivers more reliable probability forecasts for complex track variations, establishing itself as a transformative solution for operational weather prediction, especially in long-range forecasting scenarios.

From alternative perspective, end-to-end AI-driven ensemble forecasting systems have been developed (Zhong et al. 2024; Lang et al. 2024b; Chen et al. 2024; Price et al. 2025). These systems directly generate ensemble members using generative AI models. While such AI-driven ensembles, when using ERA5 reanalysis data as input, bypass the need to consider initial error effects, this approach remains impractical for operational forecasting for two key reasons: (1) real-time reanalysis data is unavailable during actual forecasting operations, and (2) forecast uncertainties would theoretically require an infinite number of members for complete characterization. Additionally, the interpretability of these ensemble members is limited by the latent-space representation of variables. Pu et al. (2025) attempted to address this by generating ensembles through initial perturbations applied to a deterministic control forecast produced by an AI model.

However, their methodology inherits similar limitations: their control forecasts depend on the unrealistic assumption of ERA5 reanalysis data availability, and the perturbations are generated using a lag-based method. Furthermore, their approach demands an impractically large ensemble size (e.g. 2000 members generated in their study) to achieve reliable forecasts. In contrast, the FuXi_CNOP system represents a significant advancement by adopting a forecasting field as initial input for control forecast generation. This doing is much applicable in realistic forecasts. Remarkably, this system achieved superior ensemble forecast skill in TC track forecasts compared to the IFS_EPS system, despite using a substantially smaller ensemble size. This innovation marks a crucial step toward operational AI-based forecasting, though further research is needed to fully realize its operational potential.

Beyond initial errors, AI forecasting systems, like numerical forecast system, encounter compounded errors from both initial conditions and model imperfections. Addressing this dual challenge, a nonlinear forcing singular vector method (C-NFSV) (Duan et al. 2022) has emerged as a breakthrough, demonstrating that dynamically coordinated initial and model perturbations in ensemble frameworks can substantially enhance forecast accuracy. When applied to AI models, this approach may not only mitigate error propagation but also unlock synergistic improvements by dynamically weighting initial and model uncertainties. In any case, for AI models exhibiting chaotic behavior, the above ensemble strategy will be particularly promising. By generating fast-growing perturbations on control forecasts generated by AI models, it could offer a transformative pathway for next-generation ensemble forecasting systems.

**Data availability**

ERA5 data is available at https://cds.climate.copernicus.eu/cdsapp#!/search?type=dataset; TIGGE data is available at https://apps.ecmwf.int/datasets/data/tigge/levtype=sfc/type=cf/; IBTrACS dataset is at https://www.ncei.noaa.gov/products/international-best-track-archive. FuXi model is open sourced at https://github.com/tpys/FuXi.

**Appendix**

To enable a quantitative evaluation, metrics for both deterministic and probabilistic forecast skills are presented. The TC track forecast error is defined as the great-circle distance between the forecasted and best-track TC centers (Cox et al. 2018), which is expressed by Eq. (A1):

$$d = 2R \cdot \arcsin\left(\sqrt{\sin^2\left(\frac{\Delta\varphi}{2}\right) + \cos\left(\varphi_1\right)\cos\left(\varphi_2\right)\sin^2\left(\frac{\Delta\lambda}{2}\right)}\right) \qquad \text{(A1)}$$

where $\Delta\varphi = \varphi_2 - \varphi_1$, $\Delta\lambda = \lambda_2 - \lambda_1$, $R = 6371$ km (Earth's radius), and $(\varphi_1, \lambda_1)$ and $(\varphi_2, \lambda_2)$ denote the latitudes and longitudes of the two points. Each TC track error is further decomposed of ATE and CTE components (Heming 2017; WMO 2013): positive CTE indicates that the forecasted track lies to the right of the best track, and positive ATE indicates that the forecasted TC moves faster than indicated by the best track.

The ensemble mean track represents the average position across all ensemble members, which is calculated by the Eq. (A2)

$$(\overline{lon}, \overline{lat}) = \frac{1}{N}\sum_{i=1}^{N}(lon_i, lat_i) \qquad \text{(A2)}$$

where $lon_i, lat_i$ are the longitude and latitude of the $i^{th}$ ensemble member, and $N$ is the number of ensemble members. The ensemble spread is defined as the standard deviation of members relative to the mean, which can be calculated by Eq. (A3):

$$\text{Spread} = \sqrt{\frac{1}{N}\sum_{i=1}^{N}|F_i - \overline{F}|^2} \qquad (A3)$$

where $F_i$ is the track of the $i^{th}$ ensemble member, $\overline{F}$ is the ensemble mean. $|F_i - \overline{F}|$ the great-circle distance (for tracks) or wind speed difference (for wind fields). A perfect ensemble forecast is proven to have a relationship in which the ensemble spread is equal to the ensemble mean forecasting error (Hopson 2014; Bowler 2006).

The CRPS measures the distance between forecasted and observed TC distributions (Gneiting and Raftery 2007; Székely and Rizzo 2013):

$$\text{CRPS} = 2E|X - Y| - E|X - X^{'}| - E|Y - Y^{'}| \qquad (A4)$$

In our research, X is the 2-dimentional TC location from the FuXi model forecast, Y is the TC location from IBTrACS data, X′ and Y′ are independent and identically distributed copies of X and Y. $E$ represent mathematical expectation and $|\bullet|$ represent represents the great-circle distance metric (physically more appropriate).

The BS is the mean squared error of the probability forecasts defined by Eq. (A5):

$$\text{BS} = \frac{1}{N}\sum_{i=1}^{N}(f_i - o_i)^2 \qquad (A5)$$

where $N$ is the number of realizations of the prediction processed, and $f_i$ and $o_i$ are the probabilities of the fore cast and observation for the prediction processes, re spectively. A smaller BS indicates a better probability forecast skill (Glenn 1950).

The ROC curve measures the ability of a forecast to dis criminate between events and nonevents. This curve is a function of the hit rate and false alarm rate. Thus, higher skill is indicated by an ROC curve that is closer to the top-left corner of the diagram (implying a low false alarm rate and high hit rate) or a larger area under the curve (ROCA) (Mason and Graham 2002).

**Acknowledgments**

This study was supported by funding from the National Natural Science Foundation of China (No. 42330111). We thank for the technical support of the National Large Scientific and Technological Infrastructure "Earth System Numerical Simulation Facility" (https://cstr.cn/ 31134.02.EL).